\documentclass[epj,nopacs,final]{svjour}
\usepackage{graphics}
\usepackage{epsfig}
\usepackage{upgreek}
\usepackage{amsmath}
\usepackage{amssymb}
\usepackage{textcomp}

\usepackage{ulem}
\begin{document}
\title{Adhesion-induced fingering instabilities in thin elastic films under strain} 
\author{Benjamin Davis-Purcell\inst{1}, Pierre Soulard\inst{2}, Thomas Salez\inst{2,3,4}, Elie Rapha\"el\inst{2} \and Kari Dalnoki-Veress \inst{1,2}
\thanks{email: dalnoki@mcmaster.ca}
}                     
\institute{Department of Physics \& Astronomy, McMaster University, Hamilton, Ontario, Canada, L8S 4M1
\and Laboratoire de Physico-Chimie Th\'{e}orique, UMR CNRS Gulliver 7083, ESPCI Paris, PSL Research University, 10 rue Vauquelin, 75005 Paris, France
\and Univ. Bordeaux, CNRS, LOMA, UMR 5798, F-33405 Talence, France
\and Global Station for Soft Matter, Global Institution for Collaborative Research and Education, Hokkaido University, Sapporo, Japan
}

\date{\today}
%
\abstract{In this study, thin elastic films supported on a rigid substrate are brought into contact with a spherical glass indenter. Upon contact, adhesive fingers emerge at the periphery of the contact patch with a characteristic wavelength.  Elastic films are also pre-strained along one axis before initiation of contact, causing the fingering pattern to become anisotropic and align with the axis along which the strain was applied. This transition from isotropic to anisotropic patterning is characterized quantitatively and a simple model is developed to understand the origin of the anisotropy.} 
%
\authorrunning{Davis-Purcell \emph{et al.}}

\maketitle
\section{Introduction}
\label{intro}
Pattern formation in thin films has been an area of great interest in soft matter physics \cite{Huang2007,Arun2009,Holmes2010,Zhou2012,Eidini2015,Leocmach2015,Paulsen2016,Cho2017}. Specifically, the study of elastocapillarity, which examines the  often competing interaction between interfacial capillary interactions and elasticity, has become a growing area of research \cite{Roman2010,Weijs2013,Brubaker2016,Schulman2017}.  Adherent materials fa\-vou\-ra\-bly contact solids, and the interfacial energy which drives the adhesion can induce a deformation that is dependent on the elastic modulus \cite{Kendall1971}. In this regard, when the adhesive is a thin film, an instability was discovered independently by Ghatak \textit{et al.} \cite{Ghatak2000} and M\"onch \textit{et al.} \cite{Monch2001} which results in stunning pattern formation. These groups found that fingering and labyrinth patterns form when adhesive contact is made with thin elastic films. This adhesion-induced fingering instability has been studied in multiple works  since the discovery~\cite{Shenoy2001,Ghatak2003,Gonuguntla2006,Ghatak2007,Vilmin2010,Chakrabarti2013,Biggins2013,Mukherjee2016,Saintyves2013,Ghosh2016,Mukherjee2017}, with much of the work summarized recently by Chaudhury, Chakrabarti and Ghatak~\cite{Chaudhury2015}.

\begin{figure}[t]
\begin{center}
\includegraphics[width = 1\columnwidth]{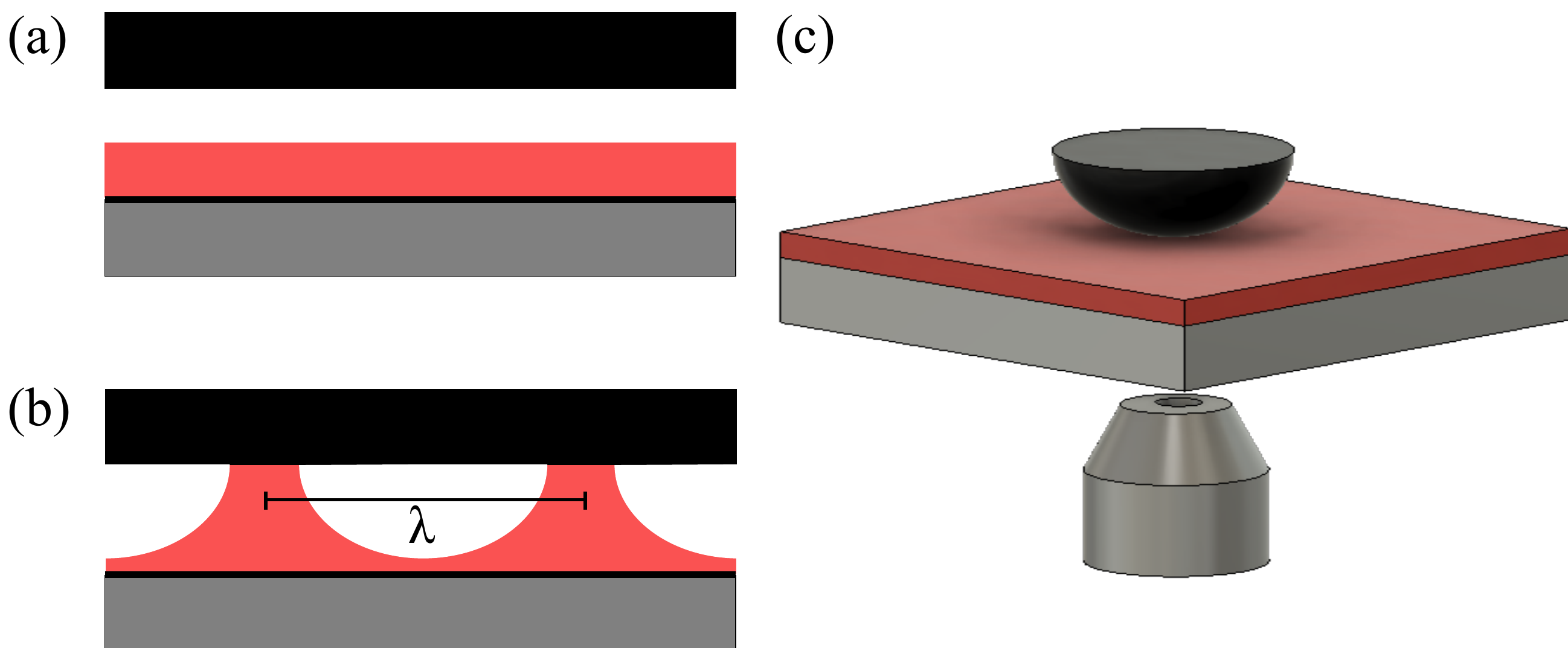}
\caption{(a) An indenter (black) is brought close to the surface of an elastic film (red) placed on a substrate (grey). (b) A characteristic wavelength, $\lambda$, forms in the elastic film when the indenter is brought into contact with the surface and contact fingers form. (c) Experimental setup: a glass spherical indenter is attached to a motorized actuator (not shown) and brought into contact with the elastic film. The contact patch is  viewed from below through a clear glass slide (shown as grey lower substrate) via optical microscopy. Schematics are not to scale.}
\label{fig:schematic}
\end{center}
\end{figure}

Experimentally, this instability may be observed when a thin elastic film is sandwiched between two rigid surfaces. An example of the instability occurs when a film is bonded to one of the rigid surfaces, while the other rigid surface -- the indenter --  is brought into contact with the free film interface. When contact is made it is seldom perfect resulting in gaps due to defects, or purposefully through the use of a spacer, or upon retraction of the indenter. This gap is crucial for the instability to develop, and experiments have been performed with spacers \cite{Ghatak2000,Monch2001,Ghatak2003}, or by ensuring complete contact is made and then minimally retracting the indenter to debond from the film \cite{Ghatak2003,Gonuguntla2006,Ghatak2007,Vilmin2010,Chakrabarti2013,Biggins2013,Mukherjee2016}. The resulting fingering pattern has a well-defined wavelength, $\lambda$, that is found to be independent of the gap formation method [shown schematically in fig.~\ref{fig:schematic}(a) and (b)].
This wavelength can be understood as a competition between interfacial energy which favours adhesion and the elastic cost associated with the deformation of the elastic film. The interfacial energy dictates that the elastic adhesive film preferentially contacts the rigid surfaces. However, in order for the adhesive film to span the gap an elastic deformation of the film is required. The last ingredient towards understanding the instability is that the volume of the elastomeric film must be conserved (Poisson's ratio for elastomers is $\nu \approx 1/2$). Thus, starting from an indenter that is some distance from the film [see  fig.~\ref{fig:schematic}(a)] the free energy can be reduced by deforming the elastomer such that contact between the indenter and the film is made. However the deformation must be accompanied by a depletion region due to volume conservation. At some distance from the contact made by the finger,  an other finger can be formed [see schematic cross-section in fig.~\ref{fig:schematic}(b)]. Starting from these simple assumptions, a theory can be developed which accurately captures the emergence of a natural wavelength \cite{Monch2001,Shenoy2001}.

In agreement with the theoretical model, numerous studies found that the wavelength depends only on the thickness, $h$, of the elastic film, and is independent of material parameters like the surface tension, $\gamma$, and shear modulus, $\mu$.  Data from many experiments were found to collapse onto one master curve well fit by $\lambda = 3.8\,h$ \cite{Chaudhury2015}. Vilmin \textit{et al.} \cite{Vilmin2010} used a simple scaling argument to show that the wavelength only depends on $h$ because of incompressibility (volume conservation), finding $\lambda \approx 4\,h$. A study by Gonuguntla \textit{et al.} \cite{Gonuguntla2006} used a linear stability analysis matched with experiments to show that for thin films where the surface tension becomes dominant in comparison to the cost of the elastic deformation, the wavelength is no longer linear with $h$. Specifically, when the  dimensionless parameter $\gamma/(\mu h)$ is large, the additional energy penalty associated with the surface tension increases the wavelength, which is especially important for thin films prepared from low modulus materials ($\mu<1$~MPa). Experiments by Chakrabarti and Chaudhury followed up on this work and expanded on the result: while still reporting a linear relationship, they found $\lambda/h > 4 $, for soft elastic materials where the elastocapillary length $\gamma/\mu$ satisfies $\gamma/\mu > h$~\cite{Chakrabarti2013}.

Multiple aspects of this adhesion-induced fingering instability have been studied since its discovery, including finger amplitude \cite{Vilmin2010} and morphology \cite{Ghatak2003}. Recent simulations and a more comprehensive theory using a cohesive-zone model have been developed \cite{Ghosh2016}. However, there are still aspects of this instability which are not fully understood, as outlined in the review by Chaudhury \emph{et al.} \cite{Chaudhury2015}. One such unknown is a quantitative analysis of the fingering instability in pre-strained films: Ghatak and Chaudhury showed qualitatively \cite{Ghatak2007} that indenting a pre-strained film causes triangular tips of the fingers along the high-strain axis, but no further study was reported.

In this article, we examine the instability pattern that appears in thin elastic films when indented with a spherical indenter, for films with $h < 1\textrm{ } \mu \textrm{m}$ and $\gamma/\mu < h$;  a regime where the linear dependence, $\lambda \propto h$, is expected to be valid. We measure the wavelength of this instability and show that the scaling of wavelength with film thickness is indeed linear, consistent with previous work. More importantly, we present experiments where the adhesive film is uniaxially strained prior to initiating contact. We find that finger growth is energetically favourable parallel to the high-strain axis. As we will show with a simple theoretical model, this is because the tension modifies the compliance of the film, which makes deformation easier in one direction relative to the other. Thus, the fingers tend to align along the high-tension axis. Lastly, we investigate how the anisotropy in the fingering pattern increases with increasing pre-strain.

\section{Experiment}
\label{expt}
Samples were prepared from a styrene-isoprene-styrene triblock copolymer (SIS, with 14\% styrene content, Sigma Aldrich) which is a physically crosslinked elastomer at room temperature. SIS films were spincoated onto freshly cleaved mica sheets (Ted Pella Inc., $2.5 \textrm{ cm} \times 2.5 \textrm{ cm}$) from solution. These solutions were made by dissolving SIS in toluene (Fisher Scientific, Optima grade) in various concentrations to make elastomeric films of varying initial thicknesses, $h_0 = \{415, 436, 510, 705, 890\}\mathrm{\ nm}$. In addition to spincoating onto mica, one sample of each thickness was spincoated onto a silicon (Si) wafer (University Wafer) for film thickness measurement using ellipsometry (Accurion, EP3). The measured film thickness on the Si was assumed to be identical to the film thickness of the films spincoated onto mica. About 20 films of each thickness were spincoated onto mica sheets. 

\begin{figure}[t]
\begin{center}
\includegraphics[width = 1\columnwidth]{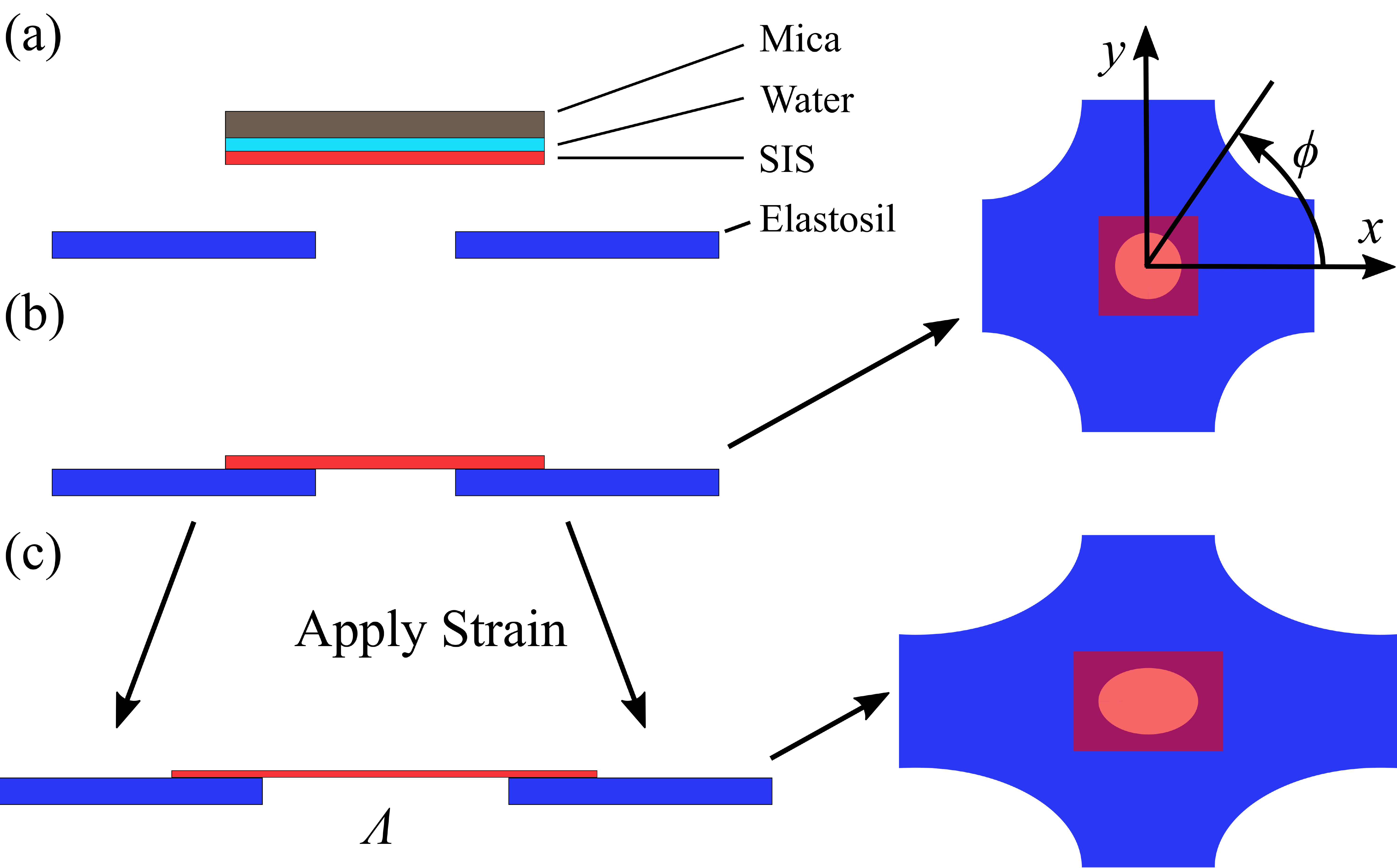}
\caption{Schematic of film transfer to custom-built stretching setup. (a) After the floating process, a thin layer of water remains between the mica sheet and the SIS film. (b) The SIS film is unstrained on the Elastosil; a corresponding top view is shown. (c) A uniaxial strain is applied, causing the hole to transition from a circle to an ellipse, uniformly straining the SIS film by a factor $\Lambda$ along the $x$-axis.}
\label{fig:stretch_setup}
\end{center}
\end{figure}

The films were annealed at $108^{\circ} \textrm{C}$ for $15 \textrm{ minutes}$ so that the triblock copolymer was in the melt state, as well as to remove excess solvent and relax the polymer chains. The SIS films were then floated onto the surface of an ultra-pure water bath ($18.2 \textrm{ M}\Omega\cdot \textrm{cm}$). The floating films were then picked back up onto the original mica sheet such that a thin layer of water between the film and the mica was present. This floating process allowed the film on mica to be transferred off the mica sheet and onto a custom-built stretching setup, which was used to apply a strain to the film along the $x$-axis while ensuring zero strain in the perpendicular $y$-axis (see fig.~\ref{fig:stretch_setup}).

The stretching setup consisted of a sheet of Elastosil with thickness $\approx200\ \mu$m (Wacker  Chemie) cut to the shape shown schematically in fig.~\ref{fig:stretch_setup}, and clamped on each of its four edges. Elastosil is an elastomer that has uniform thickness, is compliant, and remains taught when stretched. The shape of the Elastosil sheet was chosen such that upon stretching the sheet, a nearly uniform strain was applied to the central hole where the SIS film was placed.  The SIS film was transferred from the mica sheet onto the top of the Elastosil sheet, centred around the circular hole as shown schematically in fig.~\ref{fig:stretch_setup}. Pictures of the hole were taken before and after stretching and were analyzed by a custom Matlab program in order to obtain the strain conditions in the film. Films were stretched along the $x$-axis by a factor of 1 to 1.6 times their initial length, thus the extension parameter $\Lambda$ varied from 1 to 1.6, while no strain was applied along the $y$-axis (see fig.~\ref{fig:stretch_setup}). Note that the strains applied in this work were limited to $\Lambda \approx 1.6$, since beyond this point the thin free-standing films of SIS may begin to experience material failure. The limiting strain was less for the thinner films and any samples that showed signs of failure due to straining or defects were discarded. Since SIS is an elastomer with Poisson's ratio $\nu \approx 1/2$, volume is conserved (i.e. no change in density upon straining). Thus, $\Lambda_{x} = \Lambda; \Lambda_{y} = 1; \Lambda_{z} = 1/\Lambda$. As a consequence, films with an initial thickness of $h_0$ thin to a thickness of $h=h_0/\Lambda$. After the stretching process, each film was transferred to a glass microscope slide by bringing the slide into contact with the SIS film from above. SIS preferentially sticks to glass compared to Elastosil, ensuring that the strain imposed on the SIS film remained after transfer. 

\begin{figure}[h]
\begin{center}
\includegraphics[width = 0.9\columnwidth]{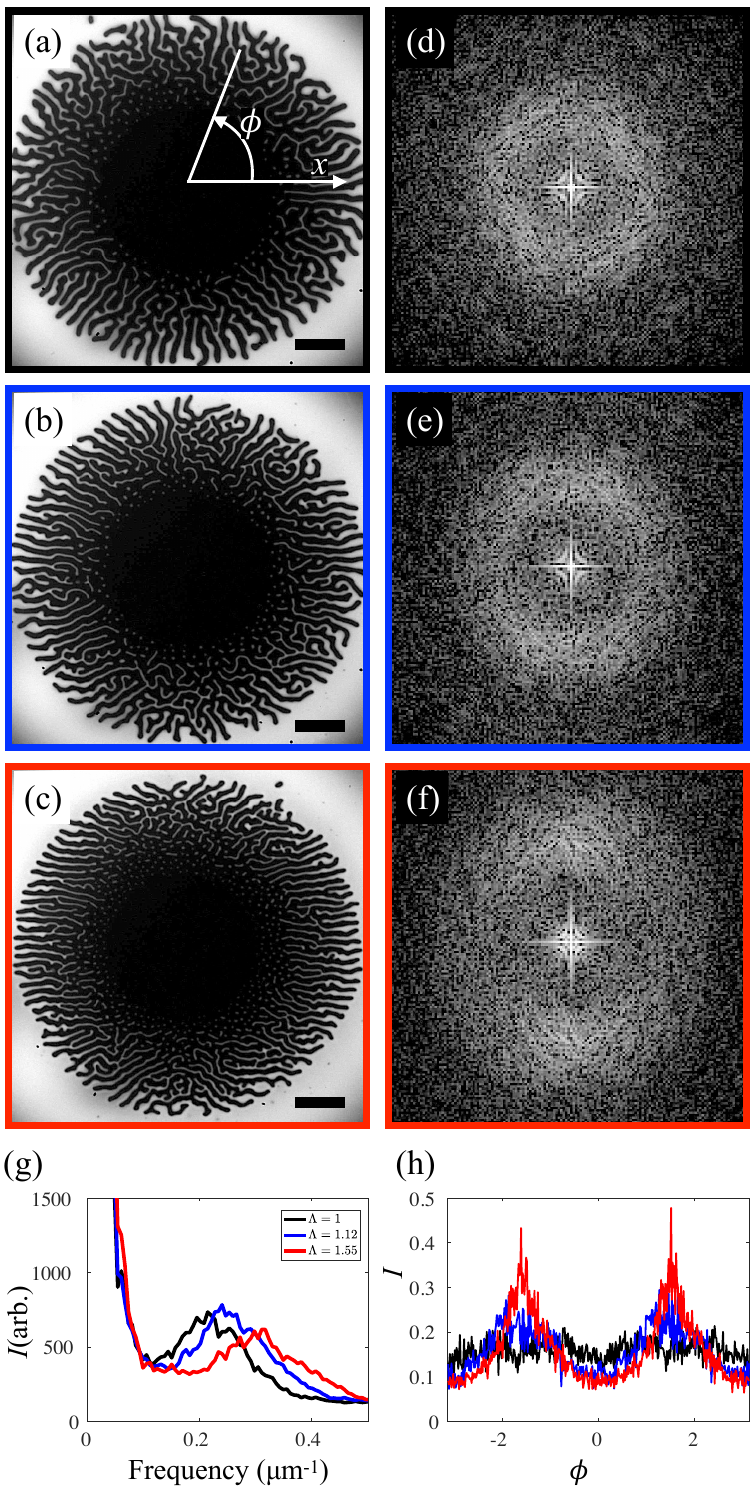}
\caption{Adhering fingers grow out from the centre of the contact patch created by the spherical indenter. For the examples shown, the films have an initial thickness $h_0=705$~nm. 
Optical microscopy images of the contact morphology are shown for different strains: a) $\Lambda=1$, b) $\Lambda=1.12$, and c) $\Lambda= 1.55$ (scale bars are $10 \textrm{ }\mu \textrm{m}$). Corresponding FFTs [(d)-(f)] are shown to the right of each image. The angular average of the Fourier intensity is shown (arbitrary units) as a function of spatial frequency in (g). The radially averaged Fourier intensity as a function of the angle is shown in (h). }
\label{fig:fft}
\end{center}
\end{figure} 

In order to carry out the adhesion contact experiment the samples were placed onto an inverted optical microscope as shown schematically in fig.~\ref{fig:schematic}(c). The indenter, a convex spherical lens (ThorLabs Inc., with a diameter of 6~mm and focal length of 30~mm) was brought into contact with the elastic film using a precision actuator (Burleigh 8200 Inchworm) until the instability pattern formed.  Images were recorded with a microscope camera (AVT GigE Vision GT1660). Sample images can be seen in fig. \ref{fig:fft}(a)-(c). 
Multiple experiments were carried out on each film: the indenter was retracted and re-positioned such that subsequent experiments were carried out on an  undisturbed area of the film. 

\section{Results and discussion}
\label{Results_and_Discussion}
The experimental data obtained in the experiments consist of the instability images -- typical results are shown in fig.~\ref{fig:fft}(a)-(c). In the images one can clearly see the central contact patch with the contact fingers radiating outwards. In the middle of the image, there is no gap and there is perfect contact between the film and the indenter, resulting in a dark contact patch due to the interference of light. As a result of the convex shape of the indenter the distance between the two rigid bodies increases away from the centre and eventually a gap opens up. Within this region the fingering instability is observed. The gap thickness increases away from the centre and the contacting fingers grow to a limiting radius where the gap is too great -- if the gap is too large, then the decrease in the free energy due to adhesion is no longer sufficient  to deform the elastic film and the fingers cannot grow.

We divide the following discussion of our results into four main parts. In section \ref{analysis}, we discuss the image analysis process. The wavelength and anisotropy parameter results are presented in section~\ref{wavelength} and \ref{order_param}, respectively. Finally, we discuss the theory in section \ref{theory}.

\subsection{Image analysis}
\label{analysis}
To analyze the data, a two-dimensional fast Fourier transform (FFT) of each image is calculated using a custom Matlab code. Sample FFTs are shown in fig.~\ref{fig:fft}(d)-(f). There are two parameters that we extract from the FFT: the wavelength, $\lambda$, of the instability pattern which corresponds to the distance between consecutive fingers, and the orientation of the fingers which we characterize by an anisotropy parameter, $s$, as defined below. 

\emph{Characteristic wavelength:} From the FFTs, we plot the angular average of the intensity as a function of spatial frequency. This plot (sample plots shown in fig. \ref{fig:fft}(g)) displays a peak, and the value of spatial frequency at the maximum of this peak corresponds to the inverse wavelength, $\lambda^{-1}$. The wavelength was calculated in such a way for each image. The measured value reported for a given film thickness is the average of three to eight images from multiple experiments on a single film, with the error calculated as the standard deviation. Results are discussed in section~\ref{wavelength}.

\emph{Anisotropy parameter:} The orientation of the pattern is encoded in the angular dependence of the Fourier intensity. The radial average is calculated over an annulus which encapsulates the ring corresponding to the high intensity; where the annulus is defined by the radii $r_{+}$ and $r_{-}$, with $r_{\pm} = \left( \lambda^{-1} \pm 0.1 \right) \mu\mathrm{m}^{-1}$. We plot the radial average of the Fourier intensity as a function of angle, $I(\phi)$, and typical data can be seen in fig.~\ref{fig:fft}(h). To quantify the orientation, we use a parameter similar to the order parameter typically used for liquid crystals \cite{Mercurieva1992,Davidson1995,Giesselmann2005,Sanchez2010}. In two dimensions, the anisotropy parameter $s$ is given by:
\begin{equation} \label{eq:s}
s = \int \left[2\cos^{2}(\phi) - 1 \right] \cdot I(\phi) \, d\phi \textrm{.}
\end{equation} 
The anisotropy parameter is calculated using equation \eqref{eq:s} independently for each image. The measured value reported for a given film thickness is the average of three to eight images from multiple experiments on a single film, with the error calculated as the standard deviation. Results are discussed in section \ref{order_param}.

\subsection{Wavelength}
\label{wavelength}
\begin{figure}[]
\begin{center}
\includegraphics[width = 1\columnwidth]{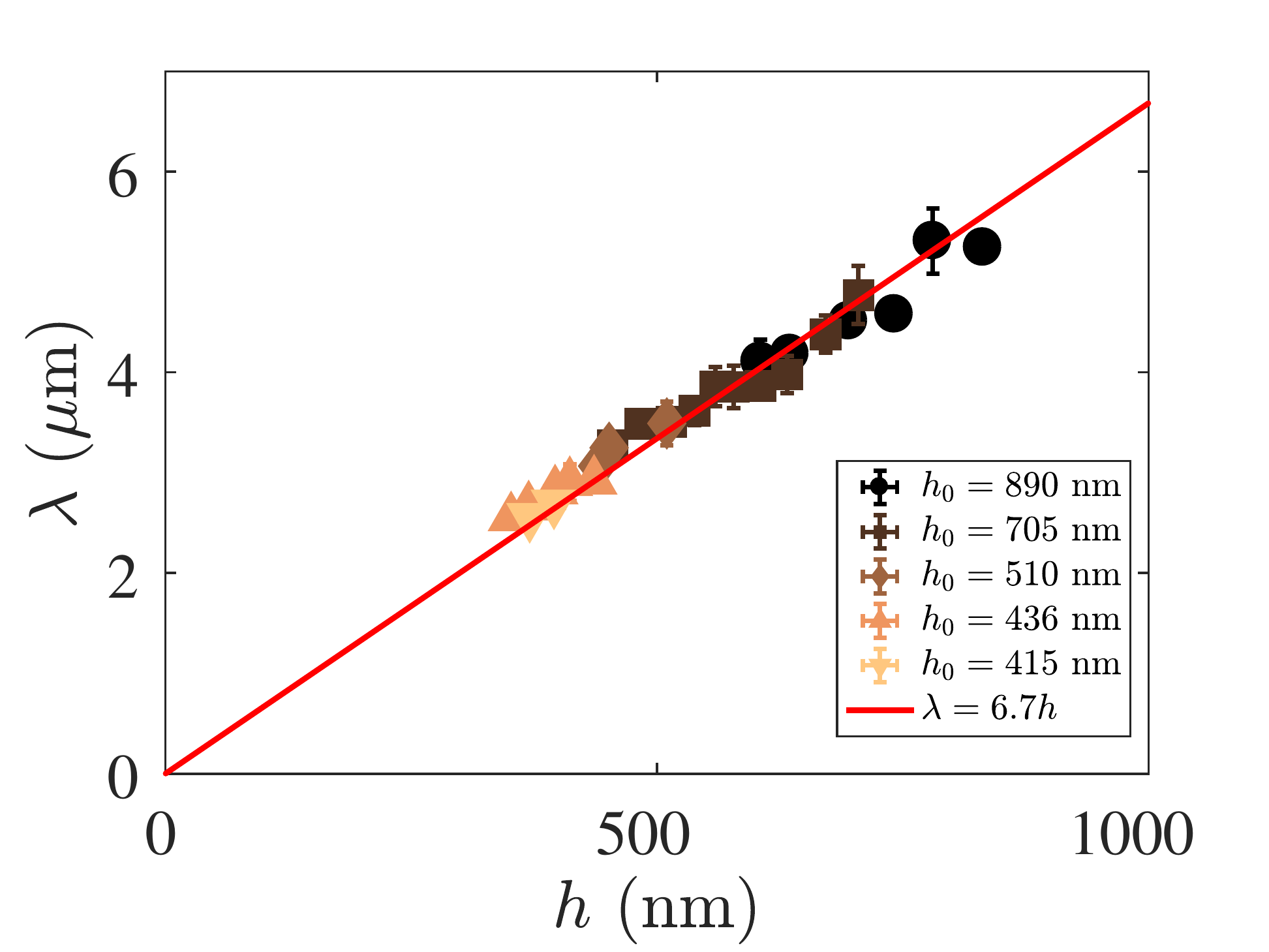}
\caption{Wavelength as a function of SIS film thickness. The data includes strained films ($1\leq\Lambda\leq1.6$), with final film thickness $h = h_{0}/\Lambda$. The data is fit to a line through the origin with a slope of $6.7$.}
\label{fig:wavelength_h}
\end{center}
\end{figure}

The wavelength is obtained from plots as shown in fig.~\ref{fig:fft}(g). The data shown in this figure corresponds to films of equal initial film thickness. As the strain is increased, the films become thinner and there is a corresponding shift in the peak towards smaller wavelengths, which is clearly visible in fig.~\ref{fig:fft}(g). Figure~\ref{fig:wavelength_h} shows the results of a full wavelength analysis. The wavelength is plotted as a function of the final film thickness (\textit{i.e.} after the strain has been imposed). The final film thickness $h$ is calculated using volume conservation as $h = h_{0}/\Lambda$. We find that all data falls on one master curve, with film thicknesses ranging from about $350 \textrm{ nm}$ to $900 \textrm{ nm}$. The wavelength is linear as a function of film thickness, which is consistent with previous work \cite{Ghatak2000,Monch2001}. The data is fit to a line forced to go through the origin as suggested by previous theoretical work and validated by experiments \cite{Shenoy2001,Vilmin2010,Chaudhury2015}. We find the slope of the fit to be $6.7$. This is a larger value than the one found by most previous work, where the relation that is most agreed upon in the literature is $\lambda = 3.8\,h$ \cite{Chaudhury2015}. Interestingly, a recent study revealed a similar prefactor to that obtained in fig.~\ref{fig:wavelength_h} with $\lambda \approx 7\,h$ \cite{Chakrabarti2013}. Therein, Chakrabarti and Chaudhury suggest that this larger slope value is due to the film material having a low shear modulus (and therefore a large elastocapillary length $\gamma/\mu$). However, the shear modulus of SIS is approximately $0.27 \textrm{ MPa}$ \cite{Schulman2017}, which is much higher than the modulus of the gel used in \cite{Chakrabarti2013}. Moreover, we can approximate the elastocapillary length $\gamma/\mu$ of SIS given its surface tension $\gamma \approx 30 \textrm{ mN/m}$ \cite{Zuo2012,Wu1970,Lee1967} and find that $\gamma/\mu \approx 110 \textrm{ nm}$. Hence, $\gamma/\mu < h$ for all film thicknesses used in this study, and the mechanism found by Chakrabarti and Chaudhury does not appear to be the cause of the relatively large slope that we find. Alternatively, we hypothesize that the large slope found in fig.~\ref{fig:wavelength_h} is a result of the specific spherical indentation geometry used in our experiments. Indeed, for a spherical indenter the fingers grow radially and the distance between fingers increases until a new finger can nucleate between them, resulting in a larger characteristic length scale in the FFT. 

\subsection{Anisotropy parameter}
\label{order_param}
We calculate the anisotropy parameter, $s$, as a function of relative strain in the film, $\Lambda$, for all films, as outlined in section~\ref{analysis}. 
\begin{figure}[]
\begin{center}
\includegraphics[width = 1\columnwidth]{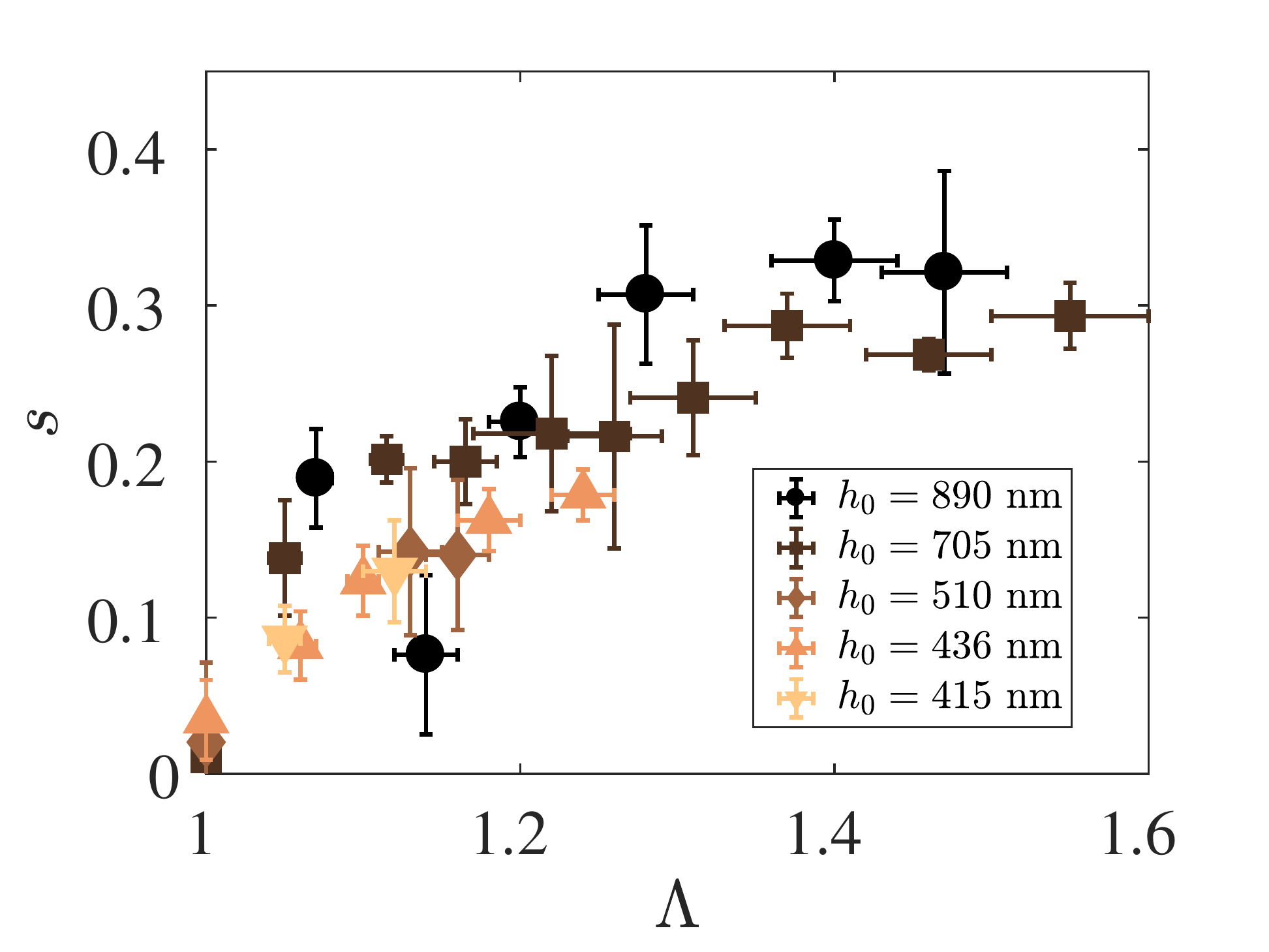}
\caption{The anisotropy parameter $s$ (see definition in eq.~(\ref{eq:s})) as a function of relative strain $\Lambda$, for several films with initial thickness $h_0$ as indicated.}
\label{fig:s}
\end{center}
\end{figure}
Figure \ref{fig:s} shows the results of the anisotropy parameter analysis. Films that are unstrained before indentation ($\Lambda = 1$) develop upon indentation a fingering pattern that is isotropic (fig.~\ref{fig:fft}(a)), \textit{i.e.} axisymmetric in the Fourier plane (fig.~\ref{fig:fft}(d)), which corresponds to $s=0$, within experimental error. Slight deviations from $s=0$ are due to pre-stresses in the film that are incurred during the transfer of the elastic films.

With increasing $\Lambda$, the fingering pattern transitions from isotropic to anisotropic, and $s$ monotonically in\-crea\-ses. Indeed, the fingers align preferentially with the high-tension axis, as seen in fig. \ref{fig:fft}(c). In addition, the anisotropy parameter appears to be insensitive to the film thickness. Note that the theoretical maximum for $s$ is 1, meaning perfect alignment of every finger along one axis. However, the experimental maximum remains below 1. This is due to the geometry of the setup and the averaging process: as the gap opens up, the fingers progress radially before bending to align with the high-tension axis. 

\begin{figure}[]
\begin{center}
\includegraphics[width = 1\columnwidth]{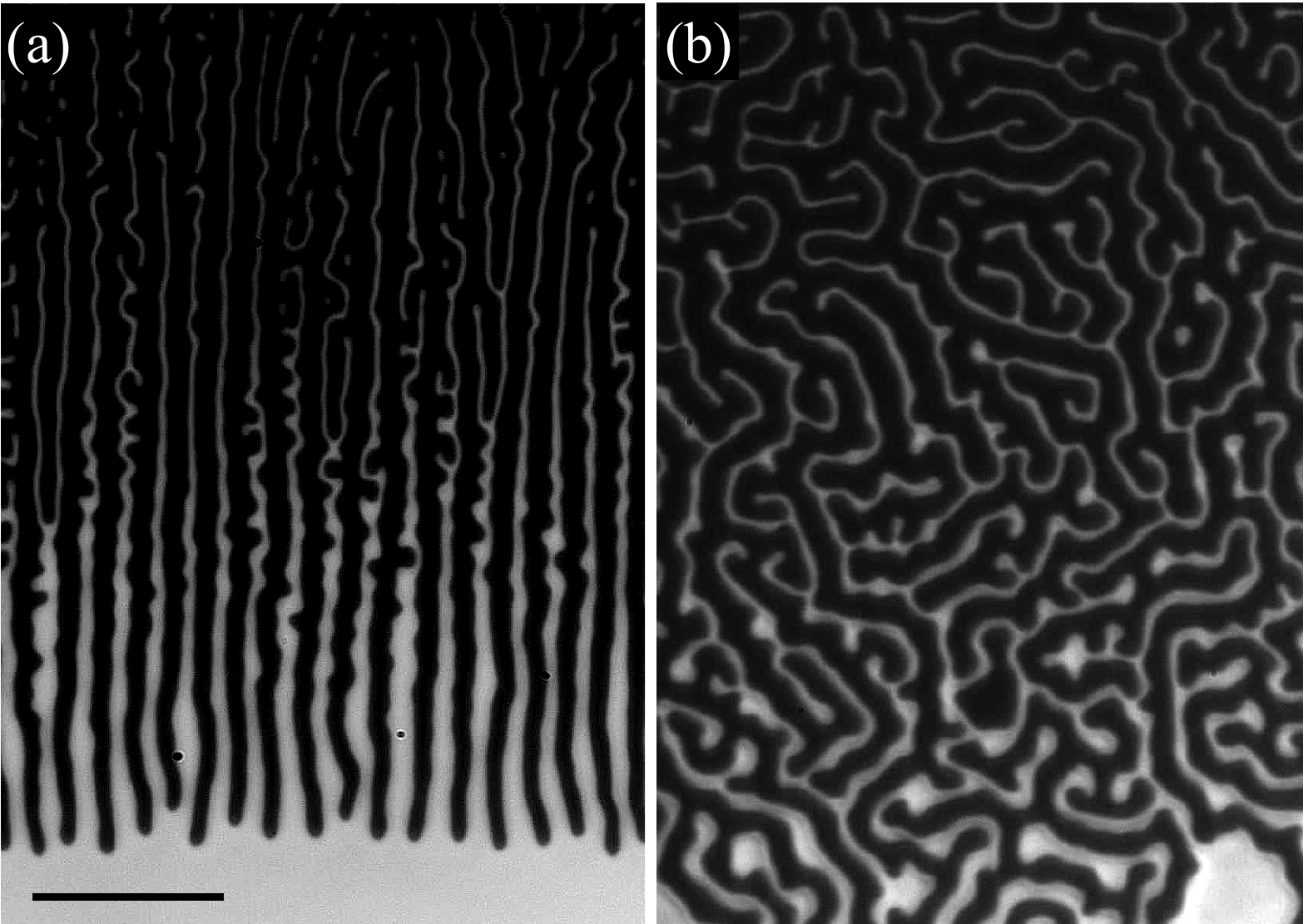}
\caption{(a) A strained SIS film ($\Lambda \approx 1.5 \textrm{; } h_{0} \approx 900 \textrm{ nm}$) indented with a cylindrical indenter. (b) Identical sample as (a) but with no applied strain. Here, the axis of the indenter is aligned along the horizontal direction and the positive strain in (a) is applied along the vertical direction. The scale bar is $20 ~\mu\mathrm{m}$.}
\label{fig:cylinder_fingers}
\end{center}
\end{figure}

In order to illustrate the anisotropy obtained by applying a pre-strain, additional experiments were carried out with a cylindrical lens on films with $h_{0} \approx 900$~nm. In fig.~\ref{fig:cylinder_fingers}(a), the morphology for a pre-strained film, with $\Lambda \approx 1.5$, is compared directly to a film with no applied pre-strain [fig.~\ref{fig:cylinder_fingers}(b)]. 

\subsection{Theory}
\label{theory}

In this section, we discuss why the fingering instability pattern aligns with the high-tension axis in the film. However, we do not discuss the wavelength scaling theory, as this has been explored extensively in previous works \cite{Monch2001,Shenoy2001,Gonuguntla2006,Vilmin2010,Chakrabarti2013,Chaudhury2015,Mukherjee2016}.

When an unstrained elastomeric film is indented with a sphere, the fingering instability pattern is axisymmetric in the Fourier plane, with an anisotropy parameter $s=0$. In contrast, under strain the fingers emerge from the contact patch radially and then re-orient axially along the $x$-direction. Qualitatively, this happens because the film is more compliant perpendicular to the high-tension direction. One can think of a stretched membrane: it is much easier to pinch the membrane perpendicular to the high tension, strained axis versus the parallel axis. Similarly here, the incompressible elastic film which is bonded to one interface must deform in order to span the gap between the rigid surfaces (see fig.~\ref{fig:schematic}(b)). Since the film is under tension along the $x$-axis, it is more compliant along the perpendicular $y$ direction, thus facilitating the deformation of the elastic surface in the $y$-$z$ plane, and therefore favouring fingers oriented along $x$.

\begin{figure}[]
\begin{center}
\includegraphics[width = 1\columnwidth]{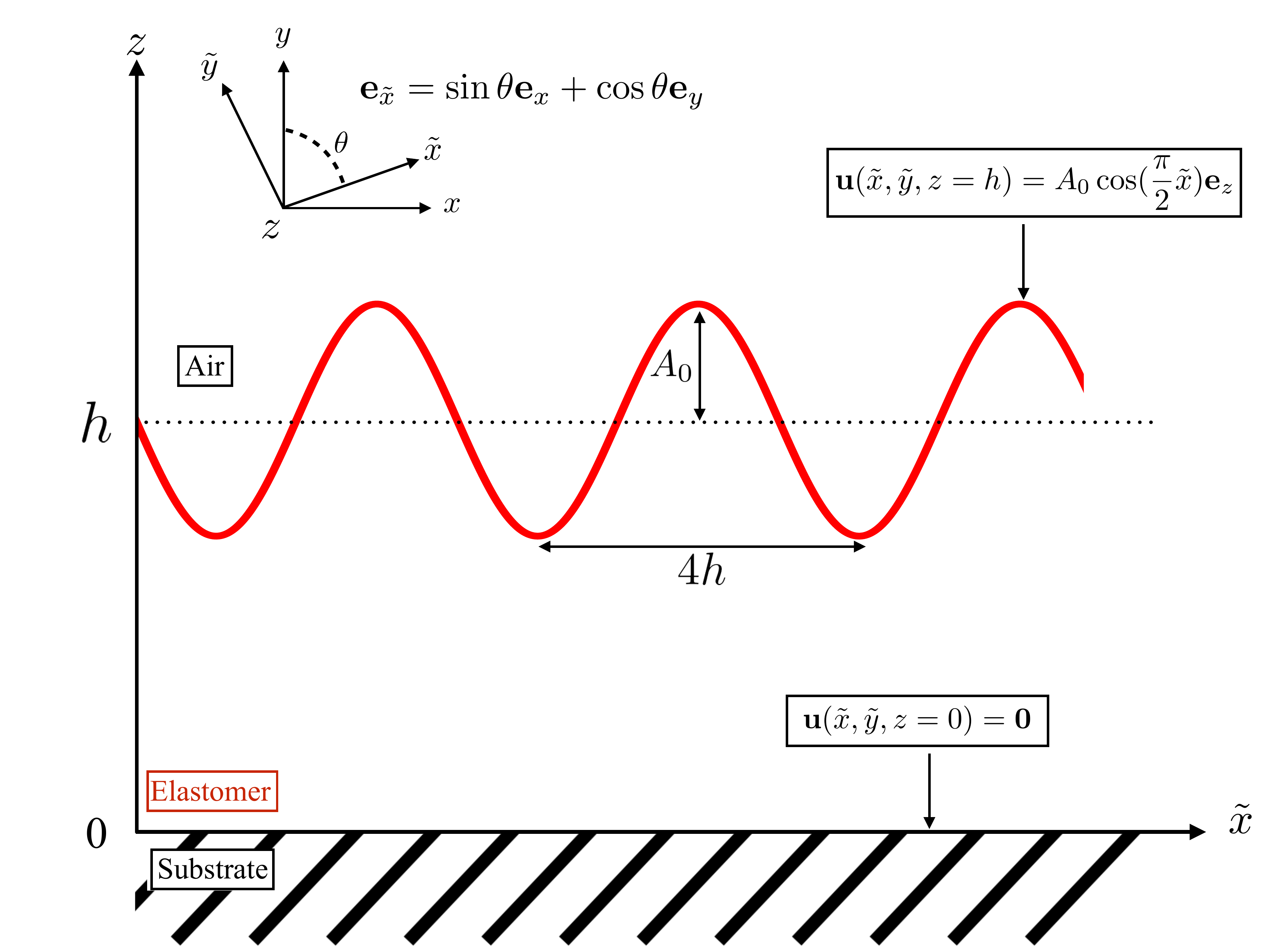}
\caption{We model the fingers by a sinusoidal perturbation of the elastomer-air interface along $z$, of amplitude $A_0$, and wavelength $\lambda=4\,h$ along an arbitrary direction $\tilde{x}$ which makes an angle $\theta$ with the $x$ stretching axis.}
\label{Fig:schema-perturbation}
\end{center}
\end{figure}

To be more quantitative, the fingers are modelled by a sinusoidal perturbation of the elastomer-air interface along $z$, of amplitude $A_0$, that is periodic along an axis $\tilde{x}$ with wavelength $\lambda=4\,h$, see fig.~\ref{Fig:schema-perturbation}. We stress that $\tilde{x}$ is arbitrary, and can in particular be misaligned with the stretching axis $x$. We thus introduce $\theta$, the angle between the $y$ and $\tilde{x}$ axes.

The following analysis is inspired by previous work \cite{Ghatak2007,Qiao2008}. We begin with a thin film in its as-prepared initial state, $O$. We then stretch the film along one axis $x$ with extension parameter $\Lambda$, to reach the stretched state $I$. The sinusoidal perturbation above (periodic along $\tilde{x}$) is then imposed as a boundary condition and results in a displacement field $\vec{u}$ in the bulk of the elastomer. We call state $II$ the final state. The strain matrices $\overline{\overline{F}}$ and $\overline{\overline{F_{0}}}$ characterize the changes from state $O$ to $II$ and from $O$ to $I$, respectively:
\begin{equation} \label{eq:matrix_transformation}
O \overset{\overline{\overline{F_{0}}}}{\rightarrow} I \overset{1+\vec{\nabla}\vec{u}}{\rightarrow} II \textrm{.}
\end{equation}
This yields: 
\begin{equation} \label{eq:F}
\overline{\overline{F}} = \left(1 + \vec{\nabla}\vec{u}\right)\overline{\overline{F_{0}}} \textrm{.}
\end{equation}
The Cauchy stress tensor is given by:
\begin{equation} \label{eq:Cauchy}
\overline{\overline{\sigma}} = 2\mu \overline{\overline{F}}: \overline{\overline{F}}^{T} \textrm{,}
\end{equation}
where $\mu$ is the shear modulus of the elastomer, and where we neglect the hydrostatic pressure. The equilibrium state $II$ is characterized by $\vec{\nabla}:\overline{\overline{\sigma}}=\bf{0}$. In addition, assuming incompressibility, one has $\vec{\nabla}.\vec{u}=0$. Using the boundary conditions indicated in fig.~\ref{Fig:schema-perturbation}, we can solve for $\vec{u}$. The elastic energy $E_{\textrm{el}}$ of a finger (per unit length in the finger's direction) is incorporated in the elastic energy (per unit length) of one period, \textit{i.e.} the trace of $\vec{\nabla}\vec{u}:\overline{\overline{\sigma}}$ integrated over one period:
\begin{equation}\label{eq:elastic_energy}
E_{\textrm{el}} = \int_{0}^{h} dz \int_{0}^{\lambda} dx \textrm{ }\textrm{Tr}\left( \vec{\nabla}\vec{u}:\overline{\overline{\sigma}}\right) \textrm{.}
\end{equation}
Finally, we obtain:
\begin{equation}\label{eq:energies}
E_{\textrm{el}}(\Lambda,\theta)=\frac{E_0}{2}\left[1+\Lambda^2(\Lambda^2\sin^2\theta+\cos^2\theta)\right]\, f(\Lambda,\theta),
\end{equation}
where $E_0=\pi\mu A_0^2/2$, and: 
\begin{equation}\label{eq:energies_final}
f(\Lambda,\theta)= \coth \left(\frac{2\pi}{\lambda} h \Lambda\sqrt{\Lambda^2\sin^2\theta+\cos^2\theta}\right). 
\end{equation}

\begin{figure}[]
\begin{center}
\includegraphics[width = 1\columnwidth]{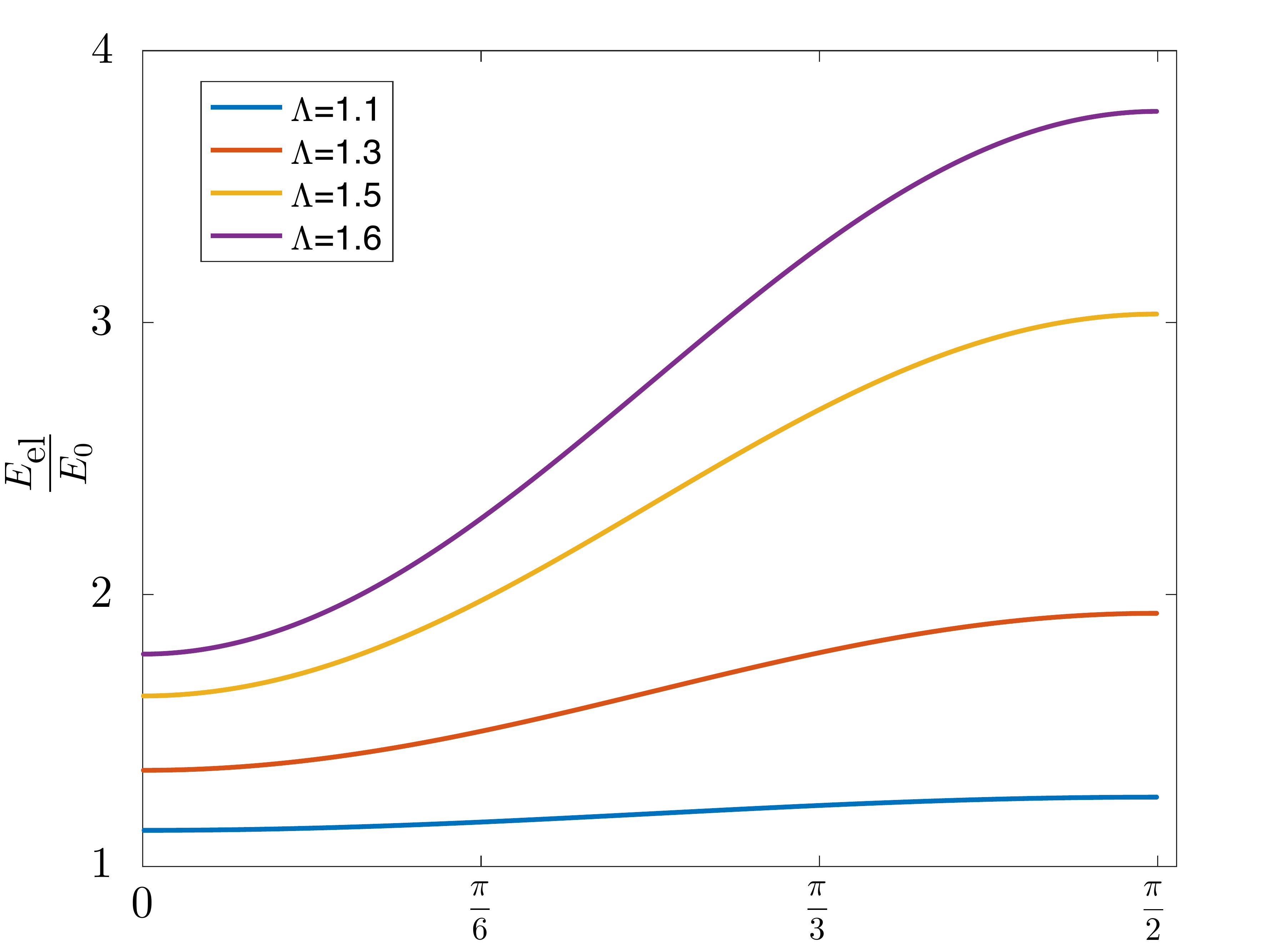}
\caption{Normalized elastic energy of a model finger (see fig.~\ref{Fig:schema-perturbation}) as a function of the orientation angle $\theta$, as given by equation \eqref{eq:energies_final}, for four values of the extension parameter $\Lambda$ as indicated.}
\label{Fig_elastic-energy}
\end{center}
\end{figure}

The elastic energy we obtained has two dependencies: the extension parameter $\Lambda$, and the orientation of the finger encoded in $\theta$. Note that, since the adhesive contribution is independent of $\theta$, we can forget it for our present matter and focus only on the elastic contribution. In fig.~\ref{Fig_elastic-energy}, we thus plot the elastic energy as a function of $\theta$ for several $\Lambda$. When $\Lambda=1$, which corresponds to no stretching, $E_{\textrm{el}} =E_0 \coth (\pi/2)$. This value is independent of the orientation of the finger, resulting in what we refer to as an isotropic morphology. In contrast, as soon as the elastomer is stretched, one has $\Lambda>1$ and the elastic energy increases with $\theta$, for $\theta$ between $0$ and $\pi/2$. Moreover, the effect is amplified with increasing $\Lambda$. This result implies that the larger the strain, the more $\theta=0$ is a favoured orientation, and thus the overall morphology becomes more anisotropic, in agreement with experiments.

Finally, we note that the preference for deformation along the low-strain direction in comparison to the high-strain direction bears similarity with the results by Chateauminois \emph{et al.},  who studied friction on elastomeric surfaces with a pre-strain~\cite{Chateauminois}. In these recent results, the authors found that the friction increases as a function of pre-strain, but that the increased friction is isotropic and independent of the indenter's geometry.

\section{Conclusion}
\label{conclusion}
In this work, thin elastic films were indented using a spherical indenter, and the resulting adhesion-induced fingering instability was observed. The wavelength of the fingering pattern was measured and was found to be in agreement with the literature. Furthermore, the orientation of the fingering pattern was characterized quantitatively as a function of pre-strain in the film. It was found that the fingers align with the axis of elongation of the elastomer. A liquid-crystal-like anisotropy parameter was introduced in order to reveal and quantify the increase of alignment with pre-strain. Interestingly, the anisotropy parameter was shown to be mostly independent of film thickness. Finally, it was shown theoretically that it is energetically favourable for the fingers created by this instability to align with the high-tension axis, confirming results from experiment.

\begin{acknowledgement}
Financial support was provided by the National Sciences and Engineering Research Council (NSERC). 
The authors also thank the Global Station for Soft Matter, a project of Global Institution for Collaborative Research and Education at Hokkaido University, and the Joliot chair from ESPCI Paris. Finally, the authors thank Solomon Barkley and Eric Weeks for useful discussions, and Wacker Chemie AG for donating the Elastosil material.
\end{acknowledgement}

%

%


\begin{thebibliography}{10}
\bibitem{Huang2007}
J. Huang, M. Juszkiewicz, W.H. de Jeu, E. Cerda, T. Emrick, N. Menon, T.P. Russell. 
Science \textbf{317}, 650–653. (doi:10.1126/science.1144616). (2007).
\bibitem{Arun2009}
N. Arun, A. Sharma, P. S. G. Pattader, I. Banerjee, H. M. Dixit, K. S. Narayan.
Phys. Rev. Lett. \textbf{102}, 254502 (2009). 
\bibitem{Holmes2010}
D. P. Holmes and A. J. Crosby.
Phys. Rev. Lett. \textbf{105}, 038303 (2010).
\bibitem{Zhou2012}
P. Zhou, S. Wise, X. Li, J. Lowengrub.
Phys. Rev. E \textbf{85}, 061605 (2012).
\bibitem{Eidini2015}
M. Eidini and G. H. Paulino.
Sci. Adv. \textbf{1} 8, 1500224 (2015).
\bibitem{Leocmach2015}
M. Leocmach, M. Nespoulous, S. Manneville, T. Gibaud.
Sci. Adv. \textbf{1} 9, 1500608 (2015).
\bibitem{Paulsen2016}
J. D. Paulsen, E. Hohlfeld, H. King, J. Huang, Z. Qiu, T. Russel, N. Menon, D. Vella, B. Davidovitch.
PNAS \textbf{113} (5) 1144-1149, (2016).
\bibitem{Cho2017}
M. R. Cho, J. H. Jung, M. k. Seo, S. U. Cho, Y. D. Kim, J. H. Lee, Y. S. Kim, P. Kim, J. Hone, J. Ihm, Y. D. Park.
Sci. Rep. \textbf{7} 43400, (2017). 
\bibitem{Roman2010}
B. Roman and J. Bico.
Journal of Physics: Condensed Matter, Volume 22, Number 49 (2010).
\bibitem{Weijs2013}
J. Weijs, B. Andreotti and J. Snoeijer. 
Soft Matter, \textbf{9}, 8494-8503 (2013).
\bibitem{Brubaker2016}
N. D. Brubaker and J. Lega.
Phil. Trans. R. Soc. A \textbf{374} 20150169 (2016).
\bibitem{Schulman2017}
R. Schulman, A. Porat, K. Charlesworth, A. Fortais, T. Salez, E. Rapha\"el and K. Dalnoki-Veress.
Soft Matter, \textbf{13}, 720 (2017).
\bibitem{Kendall1971}
K. Kendall.
Journal of Physics D: Applied Physics, Volume 4, Number 8 (1971).
\bibitem{Ghatak2000}
A. Ghatak, M. Chaudhury, V. Shenoy, and A. Sharma.
Phys. Rev. Lett. \textbf{85}, 4329 (2000).
\bibitem{Monch2001}
W. M\"onch and S. Herminghaus.
Europhys. Lett., Volume 53, Number 4 (2001).
\bibitem{Shenoy2001}
V. Shenoy and A. Sharma.
Phys. Rev. Lett. \textbf{86}, 119 (2001).
\bibitem{Ghatak2003}
A. Ghatak and M. Chaudhury.
Langmuir 19 (7), 2621-2631 (2003).
DOI: 10.1021/la026932t
\bibitem{Gonuguntla2006}
M. Gonuguntla, A. Sharma, J. Sarkar, S. Subramanian, M. Ghosh, and V. Shenoy.
Phys. Rev. Lett. \textbf{97}, 018303 (2006).
\bibitem{Ghatak2007}
A. Ghatak and M. Chaudhury.
The Journal of Adhesion, 83:679–704, (2007).
\bibitem{Vilmin2010}
T. Vilmin, F. Ziebert, E. Rapha\"el.
Langmuir, 26 (5), pp 3257–3260 (2010).
\bibitem{Chakrabarti2013}
A. Chakrabarti and M. Chaudhury.
Langmuir, 29 (23), pp 6926–6935 (2013).
\bibitem{Biggins2013}
J. Biggins, B. Saintyves, Z. Wei, E. Bouchaud, and L. Mahadevan.
PNAS 2013 110 (31) 12545-12548; doi:10.1073/pnas.1302269110 (2013).

\bibitem{Mukherjee2016}
B. Mukherjee, D. Dillard, R. Moore, R. Batra.
International Journal of Adhesion and Adhesives, Volume 66, Pages 114-127 (2016).
\bibitem{Saintyves2013}
B. Saintyves, O. Dauchot, E. Bouchaud, Phys. Rev. Lett. 111, 047801; doi:10.1103/PhysRevLett.111.047801 (2013)
%
\bibitem{Ghosh2016}
A. Ghosh, D. Bandyopadhyay, A. Sharma.
Journal of Colloid and Interface Science, Volume 477, Pages 109-122 (2016).
\bibitem{Mukherjee2017}
B. Mukherjee, R. Batra, D. Dillard.
International Journal of Solids and Structures, Vol. 110-111, 385-403 (2017).
\bibitem{Chaudhury2015}
M. Chaudhury, A. Chakrabarti, and A. Ghatak.
Eur. Phys. J. E \textbf{38}: 82 (2015).
\bibitem{Mercurieva1992}
A. Mercurieva and T. Birshtein.
Makromol. Chem., Theory Simul. \textbf{1}, 205-214 (1992).
\bibitem{Davidson1995}
P. Davidson, D. Petermann, and A. M. Levelut.
J. Phys. II \textbf{5}, 113 (1995).
\bibitem{Giesselmann2005}
F. Giesselmann, R. Germer, and A. Saipa.
J. Chem. Phys. \textbf{123}, 034906 (2005).
\bibitem{Sanchez2010}
A. Sanchez-Castillo, M. A. Osipov, and F. Giesselmann.
Phys. Rev. E \textbf{81}, 021707 (2010).
\bibitem{Zuo2012}
B. Zuo, F. F. Zheng, Y. R. Zhao, T. Chen, Z. H. Yan, H. Ni, and X. Wang.
Langmuir \textbf{28}, 4283 (2012).
\bibitem{Wu1970}
S. Wu. J. Phys. Chem. \textbf{74}, 632 (1970).
\bibitem{Lee1967}
L.-H. Lee. J. Polym. Sci. Pol. Phys. \textbf{5}, 1103 (1967).
\bibitem{Qiao2008}
L. Qiao and L. H. He.
Eur. Phys. J. E \textbf{26}, 387-393 (2008).
\bibitem{Chateauminois}
A. Chateauminois,  D. T. Nguyen, and C. Fretigny,
Soft Matter, \textbf{35}, 5849-5857 (2017).


\end{thebibliography}
 
%

\end{document}